\documentclass[12pt,a4paper]{article}
\usepackage{graphicx,hyperref}
\newcommand{\newc}{\newcommand}
\def\u#1{\verb!#1!\endgroup}

\newc{\HW}{\textsf{HERWIG}}
\newc{\TAUOLA}{\textsf{TAUOLA}}
\newc{\ThePEG}{\textsf{ThePEG}}
\newc{\HWPP}{\textsf{Herwig++}}
\newc{\evt}{\textsf{EvtGen}}
\newc{\fortran}{\textsf{FORTRAN}}
\newc{\decayer}{\textsf{Decayer}}

\newc{\HWPPClass}[1]{\href{http://projects.hepforge.org/herwig/doxygen/classHerwig_1_1#1.html}{\textsf{#1}}}
\newc{\ThePEGClass}[1]{\href{http://projects.hepforge.org/thepeg/doxygen/classThePEG_1_1#1.html}{\textsf{#1}}}
\newc{\HWPPParameter}[2]{\href{http://projects.hepforge.org/herwig/doxygen/#1Interfaces.html\##2}{{\bf #2}}}
\newc{\ThePEGParameter}[2]{\href{http://projects.hepforge.org/thepeg/doxygen/#1Interfaces.html\##2}{{\bf #2}}}
\newc{\HWPPParameterValue}[3]{\href{http://projects.hepforge.org/herwig/doxygen/#1Interfaces.html\##2}{{\bf [#2=#3]}}}
\newc{\ThePEGParameterValue}[3]{\href{http://projects.hepforge.org/thepeg/doxygen/#1Interfaces.html\##2}{{\bf [#2=#3]}}}

\begin{document}
\tolerance=100000
\thispagestyle{empty}
\setcounter{page}{0}
 \begin{flushright}
Cavendish-HEP-08/05\\
CERN-PH-TH/2008-080\\
CP3-08-08\\
IPPP/08/26\\
DCPT/08/52\\
KA-TP-09-2008\\
April 2008
\end{flushright}
\begin{center}
{\Large \bf Herwig++ 2.2 Release Note}\\[0.7cm]

M.~B\"ahr$^1$,
S.~Gieseke$^1$,
M.~Gigg$^2$,
D.~Grellscheid$^2$,
K.~Hamilton$^3$,
O.~Latunde-Dada$^4$,
S.~Pl\"atzer$^1$,
P.~Richardson$^{2,5}$,
M.~H.~Seymour$^{5,6}$,
A.~Sherstnev$^{4}$,
B.~R.~Webber$^{4}$\\[1cm]

E-mail: {\tt herwig@projects.hepforge.org}\\[1cm]

$^1$\it Institut f\"ur Theoretische Physik, Universit\"at Karlsruhe.\\[0.4mm]
$^2$\it IPPP, Department of Physics, Durham University. \\[0.4mm]
$^3$\it Centre for Particle Physics and Phenomenology, Universit\'e Catholique de Louvain.\\[0.4mm] 
$^4$\it Cavendish Laboratory, University of Cambridge.\\[0.4mm]
$^5$\it Physics Department, CERN.\\[0.4mm]
$^6$\it School of Physics and Astronomy, University of Manchester.\\[0.4mm]
\end{center}

\vspace*{\fill}

\begin{abstract}{\small\noindent
    A new release of the Monte Carlo program \HWPP\ (version 2.2) is now
    available. This version includes a number of improvements including:
    matrix elements for the production of an electroweak gauge boson, 
    $W^\pm$ and $Z^0$,
    in association with a jet; several new processes for Higgs production
    in association with an electroweak gauge boson; and
    the matrix element correction for QCD radiation in Higgs production via gluon fusion.
}
\end{abstract}

\tableofcontents
\setcounter{page}{1}

\section{Introduction}

The last major public version (2.1) of \HWPP, is described in great detail
in \cite{Bahr:2008pv}. This release note therefore only lists the
changes which have been
made since the last release~(2.1). The manual has been updated to 
reflect these changes and this release note is only intended to highlight these
new features and the other minor changes made since the last version.

Please refer to \cite{Bahr:2008pv} and the present paper if
using version 2.2 of the program.

The main new features of this version are the inclusion of matrix
elements for the production of an electroweak gauge boson, 
$W^\pm$ and $Z^0$, in association with a jet in hadron-hadron collisions,
the addition of matrix
elements for the production of an electroweak gauge boson, 
$W^\pm$ and $Z^0$, in association with the Higgs boson in both lepton-lepton
and hadron-hadron collisions, and the matrix element correction for the production
of QCD radiation in Higgs production via $gg\to h^0$.
In addition a number of other changes, such as the inclusion of 
the option of a saturation model for the small-$x$ PDF and a restructuring of
the library structure, have been
made and a number of bugs have been fixed.

\subsection{Availability}
The new program, together  with other useful files and information,
can be obtained from the following web site:
\begin{quote}\tt
       \href{http://hepforge.cedar.ac.uk/herwig/}{http://hepforge.cedar.ac.uk/herwig/}
\end{quote}
  In order to improve our response to user queries, all problems and requests for
  user support should be reported via the bug tracker on our wiki. Requests for an
  account to submit tickets and modify the wiki should be sent to 
  {\tt herwig@projects.hepforge.org}.

  \HWPP\ is released under the GNU General Public License (GPL) version 2 and 
  the MCnet guidelines for the distribution and usage of event generator software
  in an academic setting, which are distributed together with the source, and can also
  be obtained from
\begin{quote}\tt
 \href{http://www.montecarlonet.org/index.php?p=Publications/Guidelines}{http://www.montecarlonet.org/index.php?p=Publications/Guidelines}
\end{quote}

\section{New Matrix Elements}

  A number of new matrix elements are included in this release:
\begin{itemize}
\item the \textsf{MEPP2WJet} and \textsf{MEPP2ZJet} classes for the simulation
      of $W^\pm$ and $Z^0$ production in association with a hard jet
	 in hadron-hadron collisions;
\item the \textsf{MEPP2WH} and \textsf{MEPP2ZH} classes for the simulation
      of $W^\pm$ and $Z^0$ production in association with a Higgs boson
	 in hadron-hadron collisions;
\item the \textsf{MEee2ZH} class for the production of a Higgs boson
	in association with a $Z^0$ boson in $e^+e^-$ collisions.
\end{itemize}

\section{Higgs Matrix Element Correction}

\begin{figure}[!!t]
\begin{center}
\includegraphics[width=0.49\textwidth,angle=90]{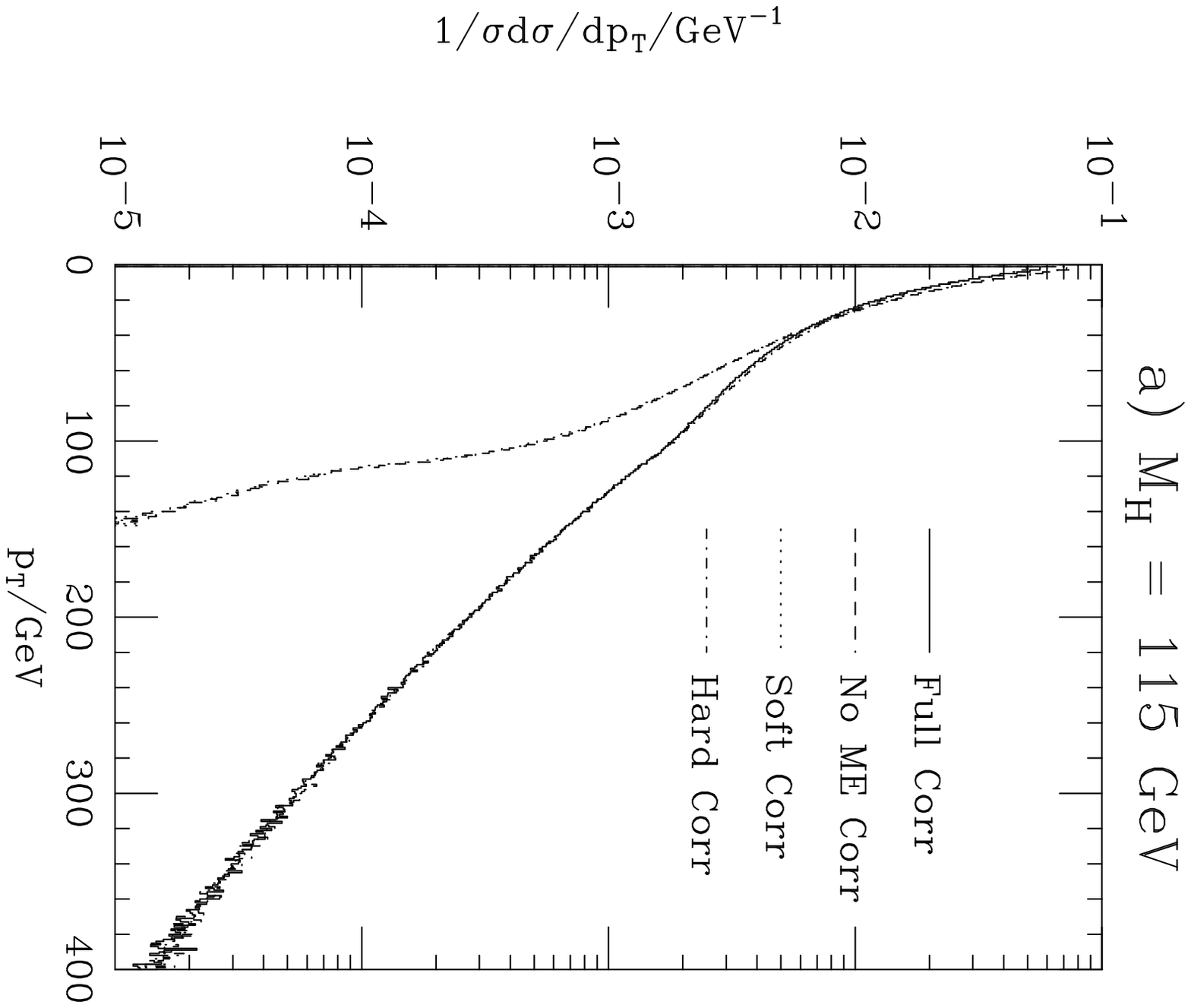}\hfill
\includegraphics[width=0.49\textwidth,angle=90]{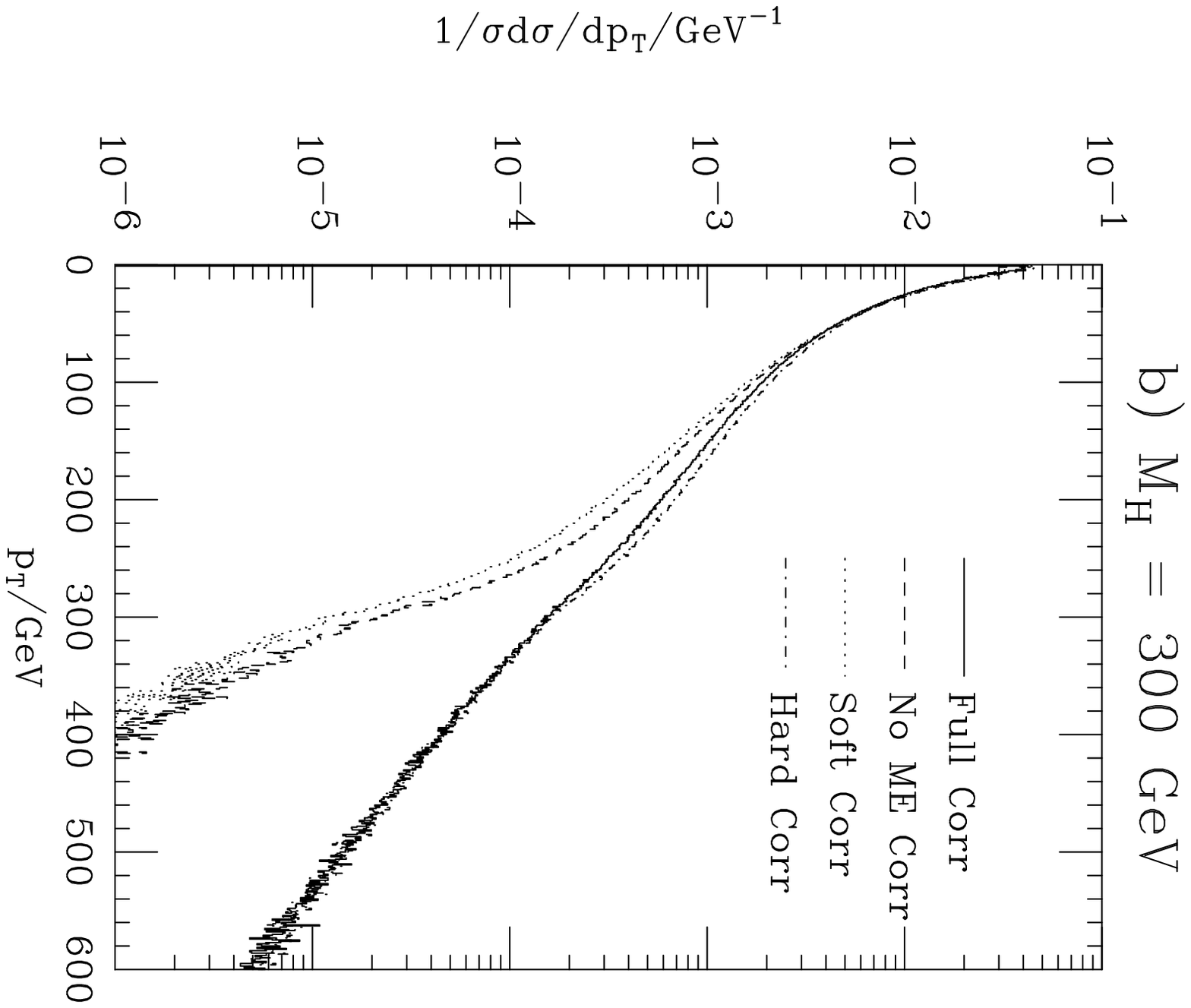}
\end{center}
\caption{The transverse momentum of the Higgs boson at the LHC for a Higgs boson
	 with mass a) 115\,GeV and b) 300\,GeV. The results of the
         full correction~(solid), no correction~(dashed), only the soft matrix element
         correction to radiation generated by the parton shower~(dotted), and the
	 hard correction filling the `dead' region not filled by the parton shower~(dot-dashed) are shown.}
\label{fig:higgspt}
\end{figure}

  The matrix element correction for $gg\to h^0$ has been included using the same approach
  as for the Drell-Yan production of a $W^\pm$ or $Z^0$ boson. The $p_T$ spectra of
  Higgs bosons with masses of 115\,GeV and 300\,GeV are shown in Fig.\,\ref{fig:higgspt}.
  Only the $gg\to h^0g$, $qg\to h^0q$ and $\bar{q}g\to h^0\bar{q}$ partonic
  processes are included in the matrix element correction as these are
  the only processes which have a parton-shower interpretation. The 
  $q\bar{q}\to h^0g$ process should therefore be included as an additional hard process
  with no cut on the $p_T$ of the outgoing gluon if all the partonic processes are
  required, in general this process gives a very small contribution.

\section{Other Changes}

A number of other more minor changes have been made.
The following changes have been made to improve the physics 
simulation:
\begin{itemize}
\item A new class implementing a saturation model for the PDFs has been included.
\item A number of additional options for \textsf{HepMC} output have been added,
      including the units to be used for energies and distances.
\item The decays of the $\Sigma_b$ baryons have been changed so that only
      the decays $\Sigma_b\to\pi\Lambda_b$ are included.
\item A number of bugs in the hadronization have been fixed necessitating
      the retuning of the
      default hadronization parameters to LEP and B-factory data. The parameters
      for the underlying event have also been retuned.
\item The option of not building the BSM models has been added, by default
      all the models are built. 
\item The particles for specific new physics models have now been moved to the files 
      specifying the models.
\item The Higgs width is now automatically recalculated from its mass and
      the limits on the off-shell mass are taken to be proportional to the width.
\item The option of generating intrinsic $p_T$ according to an inverse quadratic
      distribution has been added.
\end{itemize}

The following bugs have been fixed:
\begin{itemize}
\item A major bug in the splitting function for $g\to gg$ branchings in the initial state
      which lead to
      too little radiation being generated from incoming gluons has been fixed.
\item The scale of the veto on the production of radiation in the parton shower
      has been changed from the $p_T$ of the particles in the hard process to
      their transverse mass, this has a significant effect for top production.
\item Intrinsic $p_T$ is no longer generated for the secondary scattering
      processes generated using the multiple parton-parton scattering model.
\item The directions of hadrons containing a quark from the 
      perturbative stage of event were not correctly smeared, this has been corrected.
\item A bug in the selection of the hadrons in cluster decays which lead to $K_L^0$
      rather than $K_0$ and $\bar{K}_0$ mesons being produced has been fixed.
      A related problem which prevented more than one meson which 
      has the same flavour composition and mass being produced has also been fixed.
\item A problem with the matrix element correction for top decay in 
      rare cases where the off-shell $W$ boson mass is large has been fixed.
\item A workaround for problems with the built-in \textsf{gcc} {\tt abs} function
      has been added.
\item In order to fix some problems with the positions of particles we
      no longer include displacements for intermediates particles produced
      before the hadrons produced in the hadronization phase, all or which are
      assumed to be produced at the origin.
\item The limits on the off-shell mass of BSM particles is now set by default
      when the width is calculated or read from an input file.
\item In BSM models the branching ratios for 
      Higgs decays are now correctly reset from the SM values.
\item A number of problems related to initializing and running the generator
      using the \texttt{Herwig++ read} command have been fixed.
\item The handling of problems when the unweighting of the matrix element in the 
      generation of multiple parton-parton scattering fails has been improved.
\item Loop protection has been added to the four-body decays in the
      \textsf{Kinematics} class to prevent rare cases where an extremely large number
      of iterations were required. By default these decays now use the \textsf{MAMBO}
      algorithm to avoid this problem.
\item The value of $\pi$ used in various \textsf{AnalysisHandler}s are now
      consistently set using the value \textsf{Constants::pi}.
\end{itemize}

\section{Summary}

  \HWPP\,2.2 is the third version of the \HWPP\ program with a complete simulation of 
  hadron-hadron physics and contains incremental changes with respect to the previous
  version. The program has been extensively tested against
  a large number of observables from LEP, Tevatron and B factories.
  All the features needed for realistic studies for 
  hadron-hadron collisions are now present and  we look forward to 
  feedback and input from users, especially
  from the Tevatron and LHC experiments.

  Our next major milestone is the release of version 3.0 which will be at least as
  complete as \HW\ in all aspects of LHC and linear collider simulation.
  Following the release of \HWPP\,3.0 we expect that support for the 
  {\sf FORTRAN} program will cease.

\section*{Acknowledgements} 

This work was supported by Science and Technology Facilities Council,
formerly the Particle Physics and Astronomy Research Council, the
European Union Marie Curie Research Training Network MCnet under
contract MRTN-CT-2006-035606 and the Helmholtz--Alliance ``Physics at
the Terascale''. The research of K.~Hamilton was supported
by the Belgian Interuniversity Attraction Pole, PAI, P6/11.  M. B\"ahr
acknowledges support from the ``Promotionskolleg am Centrum f\"ur
Elementarteilchenphysik und Astroteilchenphysik CETA'' and
Landesgraduiertenf\"orderung Baden-W\"urttemberg.  S. Pl\"atzer
acknowledges support from the Landesgraduiertenf\"orderung
Baden-W\"urttemberg.

  \providecommand{\href}[2]{#2}\begingroup\raggedright\endgroup

\end{document}